# Brain Tumor Detection using Swin Transformers

Prateek A. Meshram
Computer Engineering
Dr. D. Y. Patil Institute of Engineering, Management and Research
Pune, India
prateekmeshram100@gmail.com

Suraj Sunil Joshi
Computer Engineering
Dr. D. Y. Patil Institute of Engineering, Management and Research
Pune, India
surman2825@gmail.com

Devarshi Anil Mahajan
Computer Engineering
Dr. D. Y. Patil Institute of Engineering, Management and Research
Pune, India
devarshim2002@gmail.com

*Abstract* – The first MRI scan was done in the year 1978 by researchers at EML Laboratories. As per an estimate, approximately 251,329 people died due to primary cancerous brain and CNS (Central Nervous System) Tumors in the year 2020. It has been recommended by various medical professionals that brain tumor detection at an early stage would help in saving many lives. Whenever radiologists deal with a brain MRI they try to diagnose it with the histological subtype which is quite subjective and here comes the major issue. Upon that, in developing countries like India, where there is 1 doctor for every 1151 people, the need for efficient diagnosis to help radiologists and doctors come into picture. In our approach, we aim to solve the problem using swin transformers and deep learning to detect, classify, locate and provide the size of the tumor in the particular MRI scan which would assist the doctors and radiologists in increasing their efficiency. At the end, the medics would be able to download the predictions and measures in a PDF (Portable Document Format).

## I. Introduction

A brain tumor is the growth of cells in various parts of the brain. Brain tumors can occur in various parts of the brain i.e. Cerebrum, Cerebellum and Medulla Oblongata. Over the years researchers have worked upon various techniques and technologies to step up testing rate of brain tumor detection. This area has been a prominent area of research and we have also played our part towards its betterment.

Before the starting of this project, we had also interacted with a neurosurgeon and also found out the field situations in the hospitals of India. In our study, we found that private hospitals have the best in class facilities for treating the patients. They have the requisite manpower to analyze MRI scans and provide the reports in a shorter time. But there are many patients who do not have the sufficient income available to get the treatments. So, they resort to low quality treatments. But there the problem comes that in those hospitals, there is huge workload on the medical staff to provide reports hence there are chances of delay and human error. Thus to provide a force multiplier and to better diagnose each and every section of the society we had come up with this solution.

In this proposed research, we have worked upon to develop a complete bundle of models to diagnose an MRI scan report. We have utilized a comprehensive dataset comprising around 31,354 MRI scan images which have been classified on parameters such as "Whether the scan has a tumor or not" and "What type of tumor it is".

In [1], there are various types of solutions proposed to segment the MRI scan using K-Means Clustering, ANN, SVM and so on. So, we have tried it out and worked upon to segment images using open source library OpenCV. It has been able to find a rough estimate of the size of the tumor. The main idea of implementation of this model is via the creation of REST API developed using Node.js Technology.

The motivation behind the development of such an API is to make it ready to use as a product for various healthcare functionalities and be ready for use in the market.

## II. Existing Technologies

In the world of advancements, there are several technologies through which the Detection of Brain tumors is possible. The techniques may include intelligent or non-intelligent based techniques. A few of the techniques are listed below:

A. ANN Classification:

Artificial Neural Network can be used to classify brain tumors. For this, the MRI image's textual characteristics are found using Gray Level matrix as mentioned in [2] and these features are standardized and used as input for classification using ANN.

B. SUPPORT VECTOR MACHINE:

In, [3] SVM was suggested by Hari Babu Nandpuru, Dr. S. S. Salankar, and Prof. V. R. Bora for classification of brain tumors. This method included several steps including, Pre-processing, Feature Extraction, Principle Component Analysis. This was trained on 46 images and tested on 4 images. They considered 3 kernel functions, Linear, Quadratic and polynomial and gained Sensitivity as 94.18%, 100%, and 97.06% Respectively.

C. U-NET:

P. Prakash Tunga, Vipula Singh, V. Sri Aditya, and N. Subramanya, [4] suggested U-Net: Convolutional Neural Network. U-Net has a U shaped Architecture which can be used for image segmentation. The positive predictive value for this came out 86% when trained on BRATS Dataset.

D. CA-CNN:

Wang et al. (2017) proposed Cascaded Anisotropic Convolutional Neural Network, [5] which is used for transforming multi class segmentation problem to a three layer binary segmentation problem. The Result by training on BRATS dataset resulted in 93% Sensitivity on the whole tumor.

E. CAPSNET:

In [6], Parnian Afshar, Arash Mohammadi, and Konstantinos N. Plataniotis, proposed a method of brain tumor classification using a capsule network. This led to a result of 86.56% prediction accuracy.

F. CDLLC:

In 2021, [7] Xiaoqing Gu, Zongxuan Shen, Jing Xue, Yiqing Fan and Tongguang Ni proposed a method of CDLLC that is convolutional dictionary learning with local constraints. The model was trained on the Cheng and REMBRANDT dataset and the accuracy came out to be 96.39%.

G. ANFIS:

ANFIS was proposed by P. Thirumurugan, P. Shanthakumar [8], and used the dataset from http://www.cancerimagingarchive.net/. The accuracy that they were able to achieve is 99.3%.

III. OUR PROPOSED WORK

A. DATASET DESCRIPTION

To start with, we have created our own dataset by integrating data points from various open source datasets available such as Kaggle, UCI ML, and references of datasets of various research papers available.

The dataset statistics are as follows:

| Dataset Type | Class | |
|---|---|---|
| Brain Tumor Detection Dataset | Yes | 10274 |
| | No | 5777 |
| Brain Tumor Classification Dataset | Meningioma Tumor | 10165 |
| | Glioma Tumor | 10095 |
| | Pituitary Tumor | 5317 |

**Table 1**. Dataset Details

Each of the data points has been of size 64 X 64 pixels. The Swin Transformers (Ze Liu, Yutong Lin, et. al.) was implemented to accept a batch size of 32.

To preprocess the dataset, we performed dataset augmentation, and normalization and finally converted it to tensors. The train and test split was 90 % of the training dataset and 10 % of the testing dataset. These tensors will then be plugged into the model to be developed and then trained for some epochs.

B. MODEL BUILDING

The model has been fine tuned on the pretrained Swin Transformers Model. The Swin transformers as specified in [9] works on the concept of shifted windows.

The shifted windows provide linear complexity with respect to image size. In our various experiments, we have found that compared to Vision Transformers, Swin Transformers take lower model training time and provide greater computational metrics.

In our fine tuning, we provided our own customized hyperparameters and trained it for about 3 epochs to achieve the global minimum. In our previous approaches, we have experimented on developing models using a variety of datasets and chose the best dataset. Then on those collected datasets, we trained the model and were able to achieve appreciable metrics on that comprehensive dataset. The dataset was chosen in such a way that blurry as well as clear images were included in the dataset and applied the swin transformers to work upon such data points to classify the MRI scans.

To extract features from the images, we had utilized the VitFeatureExtractor. The VitFeatureExtractor has various predefined parameters such as "sample", "image_mean", "image_std" and others. The feature extraction transformer has the ability to convert the RGB image into tensors while preserving the normalized weights for model training.

To develop the detection model, we applied a similar data collection technique. The "Yes" and "No" classes were chosen to provide the best possible metric. After then, data preprocessing to tensors, augmentation, center crop and various other techniques were applied.

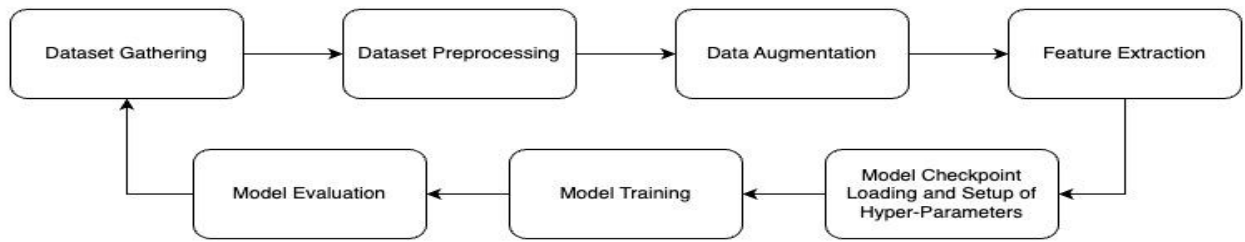

**Figure 1**. Model Development Flowchart

The classification model is able to classify the tumors based on Glioma, Meningioma and Pituitary tumors. The same above techniques are utilized to provide a good quality model. We will discuss the metrics of the models in the subsequent sections.

The segmentation of MRI scan images was performed using open source library OpenCV. In this we had tried to convert the RGB image to grayscale and shown up in the PDF report. The grayscale image was successfully able to highlight the tumor affected area in a yellow color.

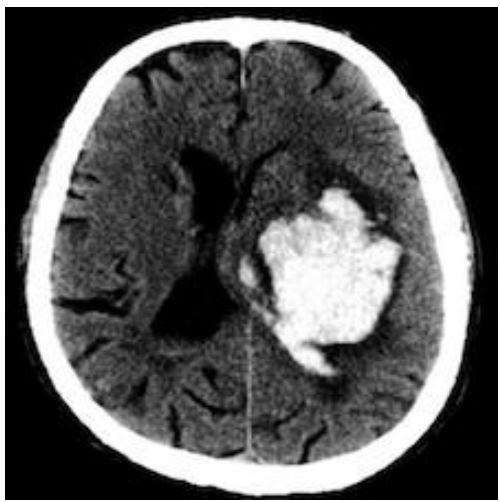 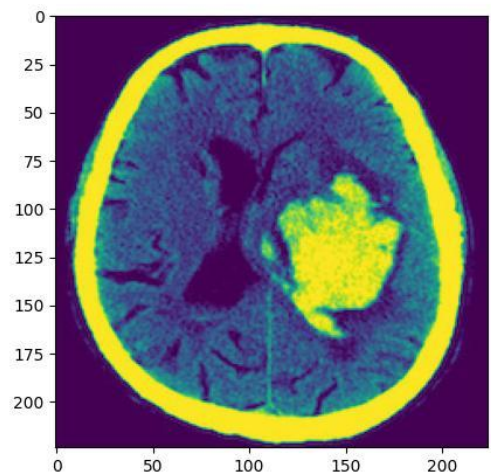

(a)                          (b)

**Figure 2**. (a) Raw image of Brain Tumor MRI, (b) Grayscale Highlighted Brain Tumor image

C. API DEVELOPMENT

The API for accessing the model and making predictions to make it ready for markets has been developed using Node.js. The API is developed in such a way that it is able to call the Python module and collect the predictions provided by the module.

The development of Node.js API has been done using various open source libraries such as child-process, PDFKit, Express and body-parser.

The API can shake hands with any healthcare software provided that it has been integrated correctly. The API would send data to and fro the python module through JSON data format owing to its light-weightedness and easy usage with a variety of programming languages.

The PDFKit library is an open source Node.js library to create PDFs in a simple way. This library was utilized to provide details about the predictions of the MRI scan. The PDF is designed in order to provide the doctors about the predictions, segmentation and size estimation of the tumor affected area.

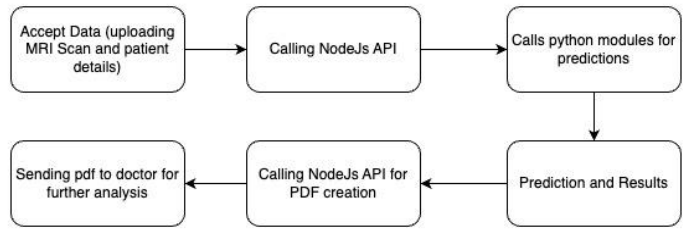

**Figure 3**. API Development Flowchart

D. EXECUTION

The following chart will be showing the execution details of the API deployment using a suitable cloud platform.

The execution of our proposed solution shows that the API would be deployed on a cloud platform which will consist of our hosted API and our trained model. The API can be called over the internet through the portals which are used as an inhouse software by the medical organizations

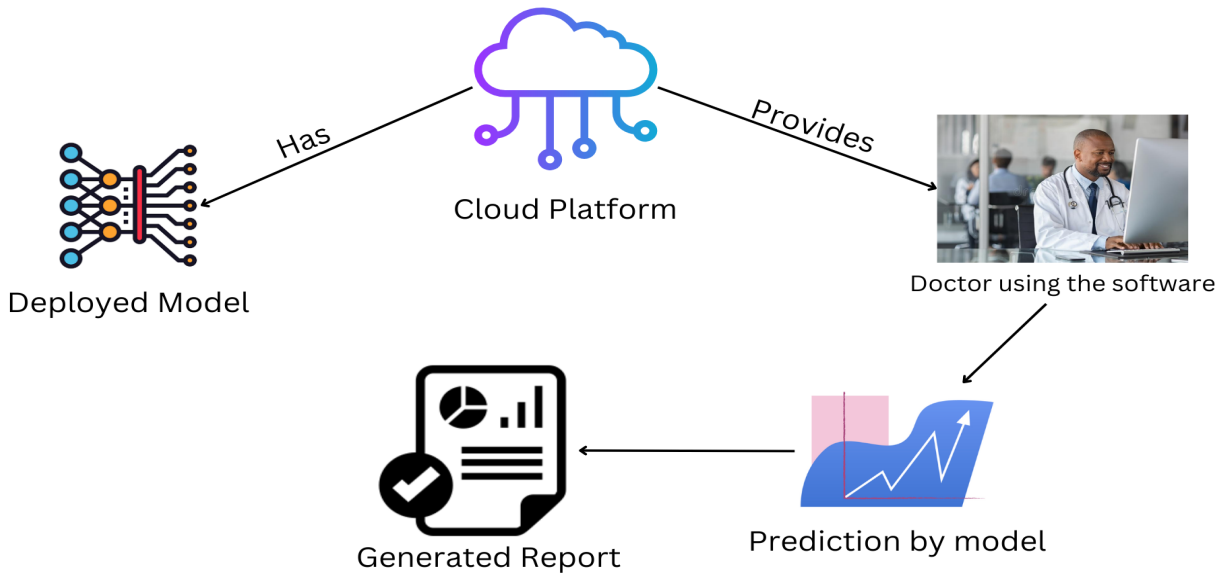

**Figure 4**. Execution of our Proposed Solution

The API calls will be made subsequently and after the predictions are done, the report will be generated and can be downloaded on the doctor's side.

## IV. Findings

### A. Performance Measure:

The performance of the proposed model for the detection of brain tumors is determined by comparing the values got by predicting the results of the test dataset (that is random 10% values of the whole dataset) to the actual values of the test dataset based on the model-trained on train dataset (that is random 90% values of the whole dataset).

The quality measures such as Sensitivity, Specificity, Fall-out, Miss Rate, PPV, NPV, F1-Score, Error Rate, and Accuracy of the proposed model of brain tumor detection were evaluated and examined. The parameters provided in the above context encapsulate the respective interpretations:

1. Sensitivity:

This measure tells how many values that are positive can the model identify correctly. The other name for sensitivity is Recall or True Positive Rate (TPR). Sensitivity or SN can be calculated by the number of correct positive predictions divided by the total number of positives.

$$Sensitivity = \frac{True\ Positive}{True\ Positive + False\ Negative}$$

2. Specificity:

This measure tells how many values that are negative can the model identify correctly. The other name for sensitivity is True Negative Rate (TNR). Specificity or SP can be calculated by the number of correct negative predictions divided by the total number of negatives.

$$Specificity = \frac{True\ Negative}{True\ Negative + False\ Positive}$$

3. Fall-Out:

This measure tells how many values that are negative can the model identify incorrectly as positive values. The other name for Fall-Out is False Positive Rate (FPR). False Positive Rate (FPR) can be calculated by the total number of negative cases incorrectly identified as positive cases divided by the total number of negative cases. It can also be calculated as 1 minus the Specificity.

$$False\ Positive\ Rate\ (Fall-Out) = 1 - Specificity$$

$$False\ Positive\ Rate\ (Fall-Out) = \frac{False\ Positive}{True\ Negative + False\ Positive}$$

4. Miss Rate:

This measure tells how many values that are positive can the model identify incorrectly as negative. The other name for Miss Rate is False Negative Rate (FNR). False Negative Rate (FNR) can be calculated by the total number of positive cases incorrectly identified as negative cases divided by the total number of positive cases. It can also be calculated as 1 minus the Sensitivity.

$$False\ Negative\ Rate\ (Miss-Rate) = 1 - Sensitivity$$

$$False\ Negative\ Rate\ (Miss-Rate) = \frac{False\ Negative}{False\ Negative + True\ Positive}$$

5. Positive Predictive Value (PPV):

This measure is the ratio of rightly identified positive cases to the total number of cases positively classified. The other name for Positive Predictive Value (PPV) is Precision.

$$Positive\ Predicted\ Value\ (Precision) = \frac{True\ Positive}{True\ Positive + False\ Positive}$$

6. Negative Predictive Value (NPV):

This measure is the ratio of rightly identified negative cases to the total number of cases negatively classified.

$$Negative\ Predicted\ Value = \frac{True\ Negative}{True\ Negative + False\ Negative}$$

7. F1-Score:

This measure tells about the accuracy of the model by considering Precision and recall. There is always a trade-off between recall and precision, hence f1 measure maximizes the trade-off. It is the harmonic mean of Precision and recall.

$$F1\ Score = \frac{Precision \times Recall}{Precision + Recall}$$

$$= \frac{True\ Positive}{True\ Positive + \frac{1}{2}(False\ Positive + False\ Negative)}$$

8. Accuracy:

   This measure tells how many times the model has predicted the results correctly. It is basically the ratio of all correct predictions divided by the total number of predictions.

$$Accuracy = \frac{True\ Positive + True\ Negative}{True\ Positive + True\ Negative + False\ Positive + False\ Negative}$$

9. Error Rate:

   This measure tells how many times the model has predicted the results incorrectly. It is basically the ratio of all incorrect predictions divided by the total number of predictions. It can also be calculated as 1 minus the Accuracy.

$$Error\ Rate = 1 - Accuracy$$

$$Error\ Rate = \frac{False\ Positive + False\ Negative}{True\ Postive + True\ Negative + False\ Positive + False\ Negative}$$

Furthermore, we compared the above values for the proposed model for brain tumor detection with the values of the other brain tumor detection technologies.

B. OUR FINDINGS:

Considering the above performance measures we calculated the performance measure of our proposed model for both models that is the Detection Model and Classification Model and have shown the results below in tabulated format:

| Sr. No. | Performance Measure | Our Proposed Model's Results | |
|---|---|---|---|
| | | Detection Model | Classification Model |
| 1 | Sensitivity / Recall / TPR | 99.90 % | 99.49 % |
| 2 | Specificity / TNR | 99.62 % | 99.786 % |
| 3 | Fall-Out / FPR | 0.38 % | 0.214 % |
| 4 | Miss Rate / FNR | 0.10 % | 0.51 % |
| 5 | PPV / Precision | 99.80 % | 99.61 % |
| 6 | NPV | 99.81 % | 99.72 % |
| 7 | F1 - Score | 99.85 % | 99.55 % |
| 8 | Accuracy | 99.81 % | 99.51 % |
| 9 | Error Rate | 0.19 % | 0.49 % |

**Table 2**. Performance Measure of Our Proposed Model

The below graphs provide an appreciable visualization of the work. The readings are taken for three epochs while training tasks. Each epoch consisted of 540 optimization steps in which the model was optimized to provide an optimal training of the model using the training dataset. The line plots are plotted for various metrics such as accuracy, precision, recall and F1 score.

The line plots provide the training metrics of both the brain tumor detection and classification models respectively.

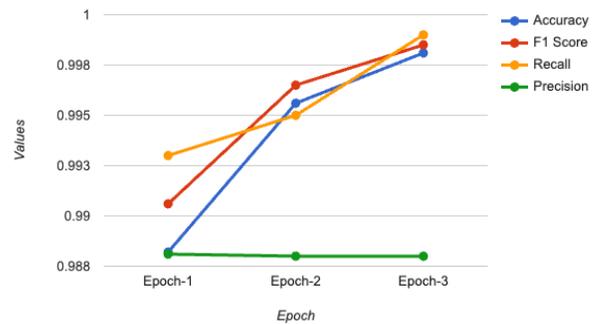

(a)

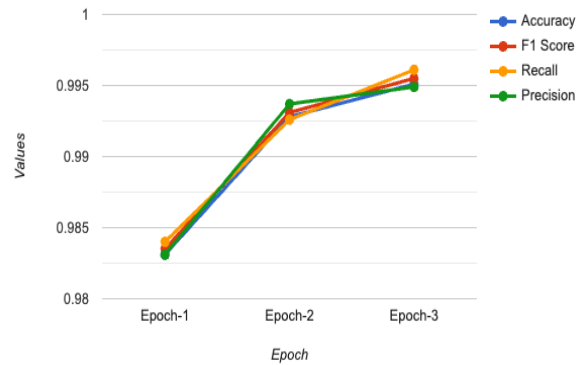

(b)

**Figure 5**. Epoch Comparison for (a) Brain Tumor Detection (b) Brain Tumor Classification

There are various Algorithms through which we can classify as well as detect brain tumors. So we compared the performance measure such as Sensitivity, Specificity and Accuracy of other algorithms for classifying brain tumor with our approach to classify the brain tumor and have displayed the results in the tabulated form as given below:

| Algorithm | Sensitivity | Specificity | Accuracy |
|---|---|---|---|
| KNN | 67 % | 83 % | 75 % |
| ELM | 90 % | 78 % | 84 % |
| FCM | 96 % | 93.3 % | 86.6 % |
| U-Net | - | - | 91 % |
| CapsNet | - | - | 92.65 % |
| SVM | 90 % | 96 % | 93 % |
| CDLLC | 94.64 % | - | 96.39 % |
| CNN | 96.4 % | 98.3 % | 97.8 % |
| ANFIS | 96.6 % | 95.3 % | 98.67 % |
| Our Approach | 99.90 % | 99.62 % | 99.81 % |

**Table 3.** Comparison of various Algorithms

To visualize the comparison between the different algorithms with our approach we plotted a Bar Graph based on the accuracy scores of all the algorithms as shown below in which blue graphs represent the accuracy scores of all other algorithms and the orange graph represent the accuracy score of our approach for brain tumor detection.

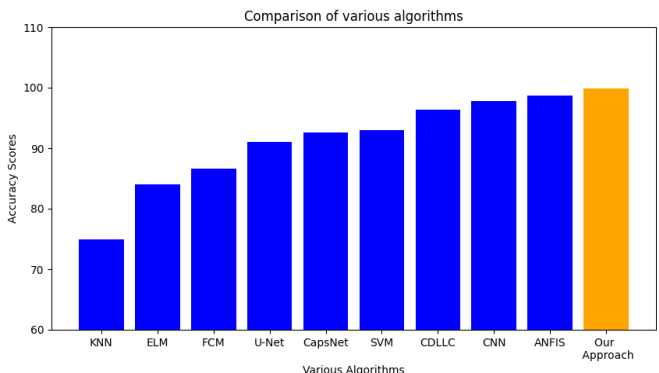

**Figure 6**. Comparison of Other Algorithms with our Approach

## IX. CONCLUSION

The current technology trend focuses on performing Brain tumor classification technique using various algorithms as specified in ANN, SVM, U-Net, CapsNet, CDLLC, ANFIS and so on, provide us the information of the various researches done in this field and the continuous advancements that were done in the recent years. So, we also researched the possibility of utilizing transformers for brain tumor detection and classification using Swin Transformers. In this, we created a comprehensive dataset comprising around 31,000 MRI scans. After performing some crucial data preprocessing steps, we came up with a solution with better metrics in which the F1 score of the model was 99% and accuracy came out to be 99.81% which is far better than the existing technologies. The solution would be implemented as an API hosted on a cloud platform through which will be able to generate a PDF report which would benefit medical professionals of varied geographical locations. The training time was about 14 hours in total. In future, we would like to implement similar such solutions for other cancer types as well and integrate them to provide a one stop solution for the doctors.